\documentclass[10pt,leqno]{amsart}
\usepackage{graphicx}
\usepackage{indentfirst}
\usepackage{tabularx} 
\usepackage{booktabs}
\usepackage{amssymb,amsthm,amsmath}
\usepackage{xcolor,paralist,hyperref,fancyhdr,etoolbox}

\hypersetup{ colorlinks=true, linkcolor=black, filecolor=black, urlcolor=black }

\usepackage{lipsum}

\begin{document}
\title[]{Clip-TTS: Contrastive Text-content and Mel-spectrogram, A High-Quality Text-to-Speech Method based on Contextual Semantic Understanding} 
\author[Initial Surname]{Tianyun Liu}
\date{\today}
\address{Address}
\email{liutianyun001@gmail.com}


\begin{abstract}
Traditional text-to-speech (TTS) methods primarily focus on establishing a mapping between phonemes and mel-spectrograms. However, during the phoneme encoding stage, there is often a lack of real mel-spectrogram auxiliary information, which results in the encoding process lacking true semantic understanding. At the same time, traditional TTS systems often struggle to balance the inference speed of the model with the quality of the synthesized speech. Methods that generate high-quality synthesized speech tend to have slower inference speeds, while faster inference methods often sacrifice speech quality. In this paper, I propose Clip-TTS, a TTS method based on the Clip architecture. This method uses the Clip framework to establish a connection between text content and real mel-spectrograms during the text encoding stage, enabling the text encoder to directly learn the true semantics of the global context, thereby ensuring the quality of the synthesized speech. In terms of model architecture, I adopt the basic structure of Transformer, which allows Clip-TTS to achieve fast inference speeds. Experimental results show that on the LJSpeech and Baker datasets, the speech generated by Clip-TTS achieves state-of-the-art MOS scores, and it also performs excellently on multi-emotion datasets. Audio samples are available at:  \href{https://ltydd1314.github.io/}{https://ltydd1314.github.io/}.
\end{abstract} 

\maketitle
\bigskip

\section{Introduction}

Speech synthesis has seen remarkable advancements in recent years, particularly with the emergence of deep learning-based TTS models. However, traditional TTS systems primarily rely on linguistic features extracted from text input, often lacking an in-depth understanding of the semantic and contextual information that influences human speech. To address these limitations, Clip-TTS introduces a novel multimodal approach by integrating Contrastive Language-Image Pretraining (Clip) [1] with text-to-speech synthesis, enabling more expressive and context-aware speech generation.  

Unlike traditional TTS models that generate speech based solely on text and phonetic information, Clip-TTS benefits from rich semantic representations, enabling it to generate speech that aligns more closely with context and human speech. By training the latent relationship between the text content and Mel spectrograms, Clip-TTS allows the text encoder to learn more information from the raw Mel spectrograms. By combining self-supervised learning and contrastive training, Clip-TTS enhances its ability to generalize across different speech styles and contexts, improving both the naturalness and adaptability of the generated speech. 

Another key advantage of Clip-TTS is its ability to control expressiveness without requiring explicit labels or large amounts of annotated data. Traditional methods often rely on manually designed prosody controls, whereas Clip-TTS implicitly learns expressive variations from multimodal data. This makes it a scalable and flexible approach, suitable for generating speech with varying tones and emotions. 

With its ability to understand and integrate text content with Mel spectrogram representations, Clip-TTS sets a new benchmark in speech synthesis. By combining multimodal learning, semantic-aware speech generation, and high-quality synthesis, it paves the way for more expressive and human-like TTS systems, significantly improving real-world applications in human-computer interaction, media production, and other fields.

\begin{itemize}
    \item I proposed a Speech-Clip framework, which establishes a connection between text content and Mel spectrograms, enabling support for multiple downstream tasks, including TTS. 
    \item Unlike conventional TTS models that rely solely on text input, Clip-TTS incorporates visual and textual context through Clip embeddings. This allows the model to capture nuanced semantic meanings and express emotions, prosody, and intonation more effectively. 
    \item By leveraging Clip's powerful semantic representations, Clip-TTS enables finer control over pitch, rhythm, and intonation, making the synthesized speech sound more dynamic and human-like. This is particularly beneficial for storytelling, character voice synthesis, and expressive speech generation in interactive applications.  
\end{itemize}

The rest of this paper is organized as follows. Section 2 mainly introduces the mainstream and classical research on text-to-speech. Section 3 focuses on Clip and its applications. Section 4 presents our proposed method. Section 6 analyzes the experimental results. Section 7 discusses the future research directions and development potential of Clip-TTS.  Conclusions are given in the final Section.

\section{Related work}

 \textbf{WaveNet:} The WaveNet series is a collection of deep learning-based speech generation models proposed by Google. These models significantly enhance the quality and naturalness of speech synthesis by directly generating waveform data. The WaveNet model is not only groundbreaking in the field of speech synthesis but has also influenced multiple related domains, such as music synthesis.
    
    WaveNet[2] is a generative model based on causal convolution that directly generates audio waveforms in the time domain. By modeling the conditional probability distribution of each sample point, WaveNet can generate speech with high fidelity and naturalness. WaveNet significantly improves the naturalness and quality of synthesized speech, with generated speech nearly reaching human-level quality. However, its generation speed is relatively slow, as it produces one sample point at a time, resulting in high computational costs and long inference times.To address the slow inference speed of WaveNet, Parallel WaveNet[3] accelerates the generation process by introducing a flow-based model. It employs a Teacher-Student Framework for training, enabling the parallel network to approximate the output distribution of WaveNet. During the generation phase, an inverse autoregressive process is used to generate waveforms in parallel, improving generation efficiency. Parallel WaveNet significantly enhances inference speed, making it capable of meeting real-time speech synthesis requirements.WaveRNN[4] further optimizes WaveNet’s computational efficiency for real-time applications. By replacing the convolutional network with a recurrent neural network (RNN) [5], it significantly reduces the number of model parameters. The advantage of this approach is a smaller model size and faster generation speed, making it suitable for resource-constrained scenarios.
   
\textbf{DeepVoice:} The DeepVoice series is a collection of end-to-end speech synthesis systems proposed by the Baidu research team. These models aim to gradually replace traditional speech synthesis pipelines through deep learning techniques, achieving high-quality and natural speech. The DeepVoice series consists of three versions, with each generation showing significant improvements in model architecture, training methods, and speech synthesis quality.

DeepVoice 1[6] built an end-to-end neural network speech synthesis system to replace traditional rule-based and statistical parameter-based TTS methods. Its advantage lies in improving the learnability between modules, enabling end-to-end training of the speech generation process. However, the naturalness of the speech generated by DeepVoice 1 was still limited, and it relied on multiple separate training modules.
DeepVoice 2[7] added support for multi-speaker speech synthesis, addressing the limitations of traditional single-speaker models. It significantly improved the model’s performance in multi-speaker speech synthesis tasks, allowing the system to quickly adapt to new speakers with minimal additional data.
DeepVoice 3[8] introduced an attention-based sequence-to-sequence (Seq2Seq) model, which greatly simplified the complexity of the speech synthesis system. The naturalness of the generated speech was significantly improved, particularly in terms of prosody and timbre, making it more realistic. It also supports a wider range of applications, including multilingual, multi-speaker, and emotional speech synthesis.

 \textbf{Tacotron:} The Tacotron series is an end-to-end speech synthesis system proposed by Google. This series of models has achieved significant advancements in speech synthesis technology, particularly in terms of naturalness, timbre diversity, and model simplification.

Tacotron[9] introduced an end-to-end model that directly generates speech spectrograms from text, replacing the complex multi-stage pipeline of traditional speech synthesis systems (such as phoneme extraction and prosody modeling). This simplification greatly reduced the complexity of conventional TTS systems and significantly improved the naturalness of generated speech, making it closer to human speech. However, Tacotron used the Griffin-Lim[10] algorithm for waveform reconstruction, which resulted in lower audio clarity. Additionally, its reliance on an attention mechanism during training could lead to alignment errors, such as syllable repetitions or omissions.
Tacotron 2[11] addressed the waveform generation issue by replacing the Griffin-Lim algorithm with WaveNet, which directly generates high-quality waveforms from spectrograms, significantly enhancing speech naturalness. Tacotron 2 became one of the most human-like speech synthesis models at the time, supporting emotional and prosody control to produce more expressive speech. However, its use of the WaveNet vocoder introduced high computational complexity, affecting real-time performance. Additionally, the model still relied heavily on large-scale training data.

 \textbf{FastSpeech:} The FastSpeech series is an efficient speech synthesis model series proposed by Zhejiang University and Microsoft Research, aimed at addressing the issues of slow generation speed, poor real-time performance, and weak prosody control in traditional end-to-end TTS models.

FastSpeech[12] improves the inference speed of TTS models and reduces the instability issues (such as attention misalignment) faced by attention-based Seq2Seq models. FastSpeech is a non-autoregressive model that bypasses the attention-based process of generating waveforms point by point. Its inference speed is more than 10 times faster than traditional autoregressive models (such as Tacotron), improving generation stability and reducing issues like misaligned or missing speech. However, its prosody and emotional expression capabilities are relatively weak, limiting the expressiveness of the generated speech.
FastSpeech 2[13] further enhances the naturalness, prosody control ability, and diversity of speech synthesis while maintaining high inference speed. It introduces more acoustic features (such as pitch, energy, etc.) as additional supervisory signals, allowing the model to control prosodic features during synthesis. It provides explicit control interfaces for adjusting speech rate, pitch, and volume. It also improves the duration predictor by incorporating alignment tools (such as Monte Carlo alignment) to achieve higher-quality phoneme duration annotations.
FastSpeech 2s[13] further shortens the generation path from text to speech, supporting direct waveform generation from text, skipping the spectrogram generation stage. This enables true end-to-end speech synthesis while maintaining high inference efficiency, achieving a better balance between inference speed and speech quality.

 \textbf{Transformer TTS:} Transformer TTS[14] is a TTS model based on the Transformer[15] architecture, proposed by Microsoft. Its core idea is to model the mapping relationship between text sequences and speech spectrograms using the self-attention mechanism while incorporating an end-to-end design to improve generation efficiency and speech quality.
Transformer TTS consists of a text encoder, a spectrogram decoder, and a vocoder. It features fast generation speed, strong stability, and support for large-scale training. Transformer TTS has been applied to various tasks, including multilingual speech synthesis, personalized speech synthesis, and emotional speech synthesis.

\textbf{NaturalSpeech:} The NaturalSpeech series is a collection of end-to-end speech synthesis models proposed by the Microsoft research team, aiming to achieve breakthroughs in generation speed, speech naturalness, diversity, and control capabilities through innovative architectures and efficient training methods.

NaturalSpeech[16] delivers high-quality speech synthesis, further enhancing the naturalness and expressiveness of speech. By adopting a non-autoregressive architecture, NaturalSpeech reduces generation steps and significantly improves speech generation speed, achieving a level of naturalness close to human speech.
NaturalSpeech 2[17] further improves the model's ability to control diverse speech features while maintaining high-quality generation, and it supports low-resource scenarios. It enhances multimodal control and data efficiency, increasing the flexibility of speech synthesis. This version can generate customized speech for specific contexts and performs better in low-resource languages or accent scenarios.

 \textbf{VITS:} VITS (Variational Inference Text-to-Speech) is an end-to-end speech synthesis model based on variational inference. The core of VITS is to combine TTS and Vocoder into a unified framework, achieving high-quality and efficient speech synthesis through end-to-end modeling.

VITS[18] introduces a parallel end-to-end text-to-speech approach, which generates audio that is more natural than current two-stage TTS models. This approach uses variational inference-enhanced normalizing flows and adversarial training to improve the expressiveness of the generative model. It also proposes a stochastic duration predictor, which can synthesize speech with different rhythms.
VITS 2[19] introduces a single-stage text-to-speech synthesis model designed to improve both the quality and efficiency of speech synthesis. It uses an adversarially trained stochastic duration predictor to synthesize more natural speech and improve efficiency. VITS 2 introduces Transformer blocks in the normalizing flow to capture long-term dependencies. It also designs a speaker-conditioned text encoder to better model speaker characteristics in multi-speaker environments.
 
\section{Clip and its applications}
 Clip [1] is a multimodal neural network proposed by OpenAI, capable of understanding the relationship between images and text, and is used for zero-shot classification, image retrieval, and other tasks.  Clip utilizes contrastive learning for joint image-text training. Its model architecture consists of an image encoder and a text encoder, which map images and texts into the same vector space, learning their associations. The training data comprises paired images and textual descriptions. Clip maximizes the similarity of matching image-text pairs while minimizing the similarity of non-matching pairs, ensuring that semantically similar images and texts are positioned closer in the vector space. After training, Clip can directly understand textual descriptions of new categories and match them with images without requiring additional training. 

AudioClip [20] is an extension of the Clip model that introduces the audio modality, enabling the model to process relationships between audio, images, and text. Its core objective is to achieve multimodal contrastive learning, aligning audio, images, and text within the same vector space to facilitate cross-modal retrieval, classification, and matching tasks.
AudioClip employs ESResNeXt [21] as its audio encoder, while its image and text encoders inherit from Clip. The training data incorporates audio, allowing the model to learn the relationships among audio, images, and text by mapping them into the same vector space during training.
After training, AudioClip can perform both bimodal and unimodal classification and querying while retaining Clip’s zero-shot inference capability. It has demonstrated excellent performance in environmental sound classification tasks, surpassing other methods.

 Clap [22] is an audio-text alignment model similar to Clip, but specifically designed for contrastive learning between audio and descriptive text. Its core objective is to map audio and text into the same vector space, ensuring that matching audio-text pairs are positioned closer while non-matching pairs are farther apart. This enables tasks such as cross-modal retrieval, classification, and zero-shot inference.
Clap's audio encoder utilizes pretrained CNNs (such as ESResNeXt), Transformers (such as AST [23], HTS-AT [24]), or HuBERT [25]/Wav2Vec [26] variants. Its text encoder is based on BERT [27], RoBERTa [28], or Clip’s text encoder. Without requiring additional training, Clap can directly understand new categories of audio and match them with textual descriptions.

ClapSpeech [29] is a cross-modal contrastive pretraining framework designed to explicitly learn the prosodic variations of the same textual tokens in different contexts. Its core idea is to use contrastive learning to associate textual contexts with their corresponding prosodic patterns.
ClapSpeech consists of a text encoder and a prosody encoder. The text encoder is trained to predict prosody from textual context, while the prosody encoder extracts real prosodic features from speech segments. This prosody encoder can be effectively integrated into TTS systems.

\section{Method}
Clip learns the matching relationships between images and their corresponding descriptive text through contrastive learning, enabling support for various downstream tasks. My inspiration is based on the idea of Clip, adopting a similar contrastive learning approach to learn the latent mapping between Mel spectrograms and corresponding text content. This allows us to train a foundational model, which I call Speech-Clip. Theoretically, this model can also support many downstream tasks.
This paper focuses on the TTS branch, while more potential branches will be discussed in Section 7.

\begin{figure}[ht]
    \centering
    \includegraphics[width=0.8\linewidth]{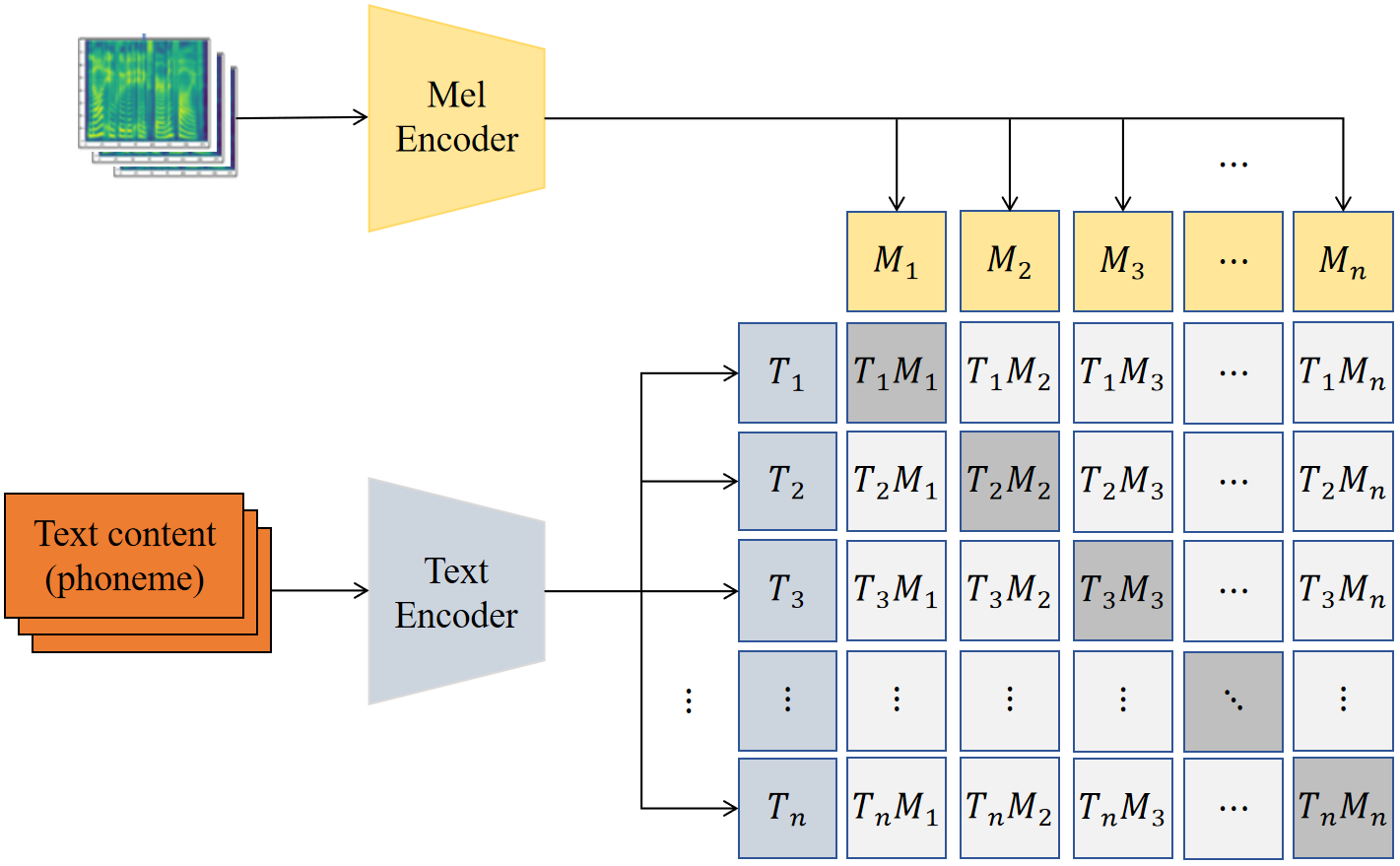}
    \caption{The general framework of Speech-Clip}
    \label{fig:1}
\end{figure}
Figure 1 shows the basic framework of our Speech-Clip, which consists of a text encoder and a Mel encoder. The text encoder takes text information as input and encodes it, while the Mel encoder receives the Mel spectrogram as input. Finally, I calculate the cosine similarity between the outputs of the two encoders. For matching text content and Mel spectrogram, I hope the cosine similarity approaches 1, and for non-matching pairs, it should approach 0. 

However, the length of the text and the length of the Mel spectrogram are never equal, which results in unequal output length from the two encoders. Cosine similarity cannot be computed between two unequal vectors. Clip addresses this issue by using a linear projection layer, which projects the outputs of both encoders to the same length, allowing for the calculation of cosine similarity. However, in the field of speech synthesis, simply using a linear projection to forcibly align the text and Mel spectrogram is not feasible, as this does not ensure phoneme-level alignment. Therefore, I encapsulate a duration predictor within the text encoder, which aligns the length of the text with the Mel spectrogram at the phoneme level, thus achieving alignment of the text and Mel spectrogram in a unified length. 

\begin{figure}[ht]
    \centering
    \includegraphics[width=0.8\linewidth]{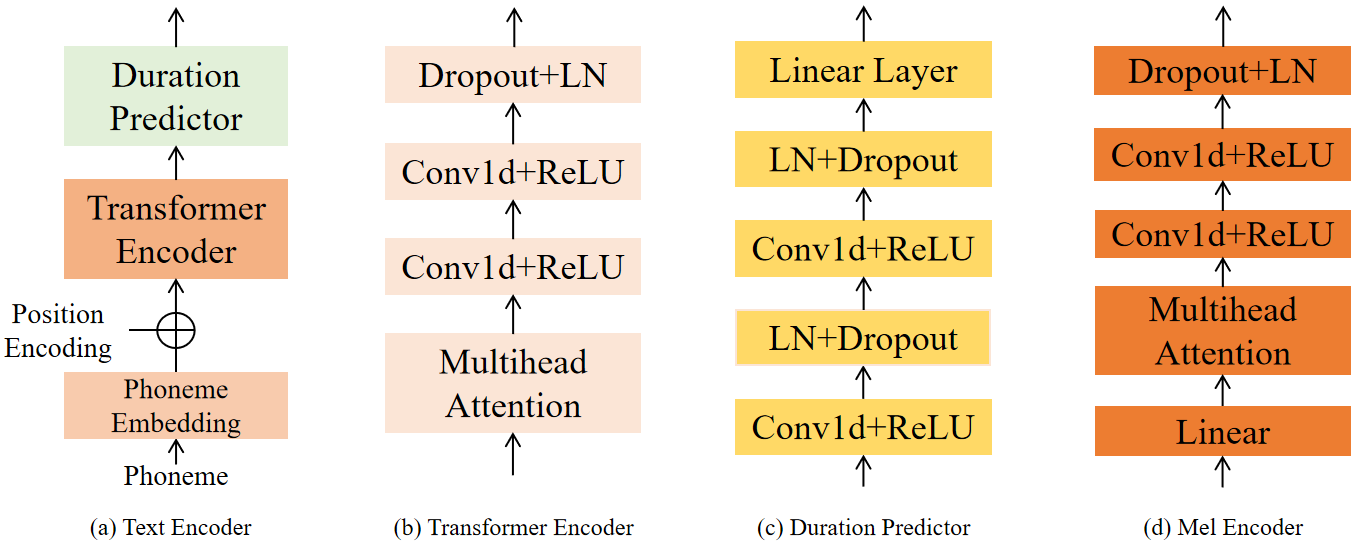}
    \caption{The general framework of Text Encoder}
    \label{fig:5}
\end{figure}

Figure 2 (a) shows the basic framework of our text encoder. The text encoder consists of three main components: phoneme embedding, transformer encoder, and duration predictor. The input to the transformer encoder is the phoneme encoding, which is generated by the phoneme embedding and is also provided with positional encoding. The output of the transformer encoder is then passed to the duration predictor for phoneme alignment.  Figure 2 (b) shows the main structure of my transformer encoder. The encoder consists of a multi-head attention, two layers of 1D convolutions with ReLU, and a layer of dropout followed by layer normalization. This structure allows us to balance both contextual information and global features, while also preventing overfitting. I directly use the structure from FastSpeech2 for the duration predictor, as shown in Figure 2 (c). Figure 2 (d) shows the structure of our Mel encoder. Unlike the text encoder, I have added a linear layer before the multi-head attention in the Mel encoder. The purpose of this layer is to increase the dimensionality of the Mel spectrogram from 80 to 256, so that it matches the dimensionality of the phoneme embedding. 

\begin{figure}[ht]
    \centering
    \includegraphics[width=0.8\linewidth]{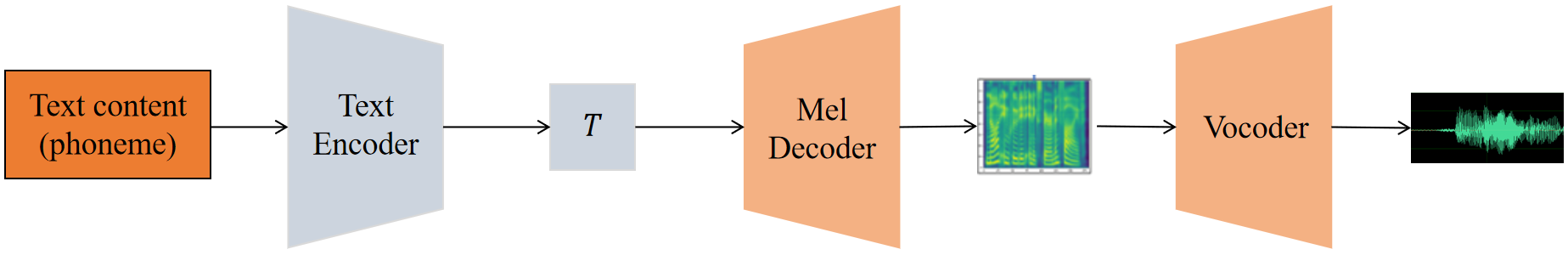}
    \caption{The general framework of Clip-TTS}
    \label{fig:2}
\end{figure}
After training the Speech-Clip model, I extract the text encoder and add a Mel decoder (with fine-tuning) and vocoder after it, thus forming a complete TTS system, as shown in Figure 3. The Mel decoder is responsible for decoding the output of the text encoder into a Mel spectrogram, while the vocoder converts the Mel spectrogram back into a waveform. The structure of the Mel decoder also adopts the transformer encoder architecture, with the difference being that the multi-head attention is replaced by the mask multi-head attention. Additionally, a linear layer is added at the end to reduce the dimensionality from 256 back to 80.    

\section{Experiment and Results}

\textbf{Datasets:} I selected two datasets for both Chinese and English, including single-speaker and multi-speaker datasets, making a total of four datasets: Baker [30], AISHELL3 [31], LJSpeech [32], and LibriTTS [33]. Based on these four datasets, I trained four models. Baker and LJSpeech are datasets recorded by a single female speaker, with 10,000 samples (about 12 hours) and 13,100 samples (about 24 hours), respectively. They are characterized by high quality, standard and clear pronunciation, wide phoneme coverage, and abundant labels, but their drawback is that they cannot meet the multi-speaker requirement. AISHELL3 and LibriTTS contain 218 speakers (about 85 hours) and 2,456 speakers (585 hours), respectively. These datasets have a large volume of data, wide phoneme coverage, and can meet the multi-speaker requirement, but they suffer from issues such as non-standard pronunciation and poor speech quality. In addition, I also experimented with a multi-emotion speech dataset, the Emotional Speech Dataset (ESD) [34]. This dataset contains ten speakers for both Chinese and English, each expressing five emotions (anger, happy, neutral, sad, and surprise), with each speaker recording 350 samples for each emotion. Therefore, I divided this dataset by language and packaged it into the Baker and LJSpeech datasets, forming a bilingual dataset with five different emotional categories. I used this dataset to train two models, one for Chinese and one for English. 

\textbf{Model Structure:} My Clip-TTS consists of 4 feed-forward Transformer (FFT) blocks in the text encoder, the Mel encoder and the Mel decoder. In each FFT block, the dimension of phoneme embeddings and the hidden size of the multi-head attention are set to 256. The number of attention heads is set to 2 and the kernel sizes of the 1D-convolution in the 2-layer convolutional network after the self-attention layer are set to 9 and 1, with input/output size of 256/1024 for the first layer and 1024/256 in the second layer. The output linear layer converts the 256-dimensional hidden states into 80-dimensional mel-spectrograms and optimized with mean absolute error (MAE). In the duration predictor, the kernel sizes of the 1D-convolution are set to 3, with input/output sizes of 256/256 for both layers and the dropout rate is set to 0.5.

\textbf{Training and Inference:}  I train my Clip-TTS on 1 NVIDIA GeForce RTX 4090 GPU, with batchsize of 16 sentences. I use the Adam optimizer [35] and follow the same learning rate schedule in [15]. It takes 90k steps for training until convergence. In the inference process, the output mel-spectrograms of my Clip-TTS are transformed into audio samples using pre-trained HiFi-GAN [36].
\begin{table}[ht]
    \centering
\caption{The MOS with 95\% confidence intervals on varies datasets}
\label{tab:1}

    \begin{tabularx}{\textwidth}{>{\centering\arraybackslash}p{5cm}>{\centering\arraybackslash}X>{\centering\arraybackslash}X>{\centering\arraybackslash}X>{\centering\arraybackslash}X} \toprule  
         &\multicolumn{4}{c}{MOS}\\   \cmidrule(lr){2-5}
 Method& LJSpeech&Baker&AISHELL3&LibriTTS\\   \midrule
         GT& 4.43±0.08& 4.82±0.09& 4.12±0.05&4.31±0.08\\   
         GT(Mel+HiFi-GAN)& 4.45±0.06& 4.72±0.04& 4.17±0.09&4.25±0.13\\ 
 CDFSE& -& -& 3.47±0.11&-\\  
         FastSpeech& 3.71±0.10& -& -&-\\   
         FastSpeech2& 4.24±0.19& -& 3.72±0.05&3.79±0.11\\   
         vTTS& 4.08±0.24& -& -&-\\   
         VAENAR-TTS& 4.33±0.07& 4.47±0.15& -&-\\   
         Clip-TTS& 4.35±0.15& 4.33±0.04& 3.80±0.10&3.75±0.24\\ \bottomrule
    \end{tabularx}

\end{table}

The speech quality evaluation method I used is the MOS [37] score. The comparative experiments were conducted mainly with the FastSpeech series and its extended works. In addition to FastSpeech 1/2, I also included Tsinghua University's CDFSE [38] and the University of Tokyo's vTTS [39]. For the Baker dataset, I chose VAENAR-TTS [40], which is currently one of the best-performing models. The experimental results are shown in Table 1. As we can see, on the LJSpeech dataset, my method outperforms others. On the Baker dataset, my method is close to the optimal level, and on both of these datasets, my method approaches the ground truth level. On the AISHELL3 dataset, my method outperforms others, while on the LibriTTS dataset, my method is slightly lower than FastSpeech 2. Since the ground truth scores for the AISHELL3 and LibriTTS datasets are not very high, my method also scored relatively low on these two datasets. This indicates that my method is notably dependent on the quality of the dataset, which may be related to my limited computational resources, as I am unable to train large models. As a result, the model's performance and robustness are somewhat below expectations. 
\begin{table}[ht]
    \centering
     \caption{The CMOS with 95\% confidence intervals on emotion datasets}
    \label{tab:2}
    \begin{tabularx}{\textwidth}{>{\centering\arraybackslash}X>{\centering\arraybackslash}X>{\centering\arraybackslash}X>{\centering\arraybackslash}X>{\centering\arraybackslash}X>{\centering\arraybackslash}X} \toprule  
         &  \multicolumn{5}{c}{CMOS}\\ \cmidrule(lr){2-6}  
         Datasets&  Neutral&  Angry&  Happy&  Sad& Surprise\\ \midrule  
 LJSpeech& +0.069& +0.045& -0.058& -0.028&-0.144\\ 
         Baker&  +0.035&  -0.125&  -0.134&  +0.057& +0.033\\ \bottomrule
    \end{tabularx}
   
\end{table}

On the emotional speech dataset, I used CMOS scores to evaluate the differences between the speech generated by Clip-TTS for different emotions and the original speech. The CMOS scores are divided into 6 levels, ranging from -3 to 3. A score between -1 and -3 indicates that the quality of the synthesized speech is progressively worse than the original speech, while a score between +1 and +3 indicates the opposite. If the similarity between the synthesized speech and the original speech is close, the CMOS score will be close to 0. As shown in Table 2, on the LJSpeech and Baker datasets, my method is able to generate speech with different emotions, and the similarity to the original speech is generally consistent.

\section{Future research directions and development potential }

The success of Clip-TTS proves the effectiveness of the Clip architecture in the field of speech synthesis. Currently, I have completed the development of a basic TTS system, but the potential of Clip-TTS and Speech-Clip goes beyond that. In this section, I will focus on discussing the further development of Clip-TTS and the potential of Speech-Clip. 

\begin{figure}[ht]
    \centering
    \includegraphics[width=0.8\linewidth]{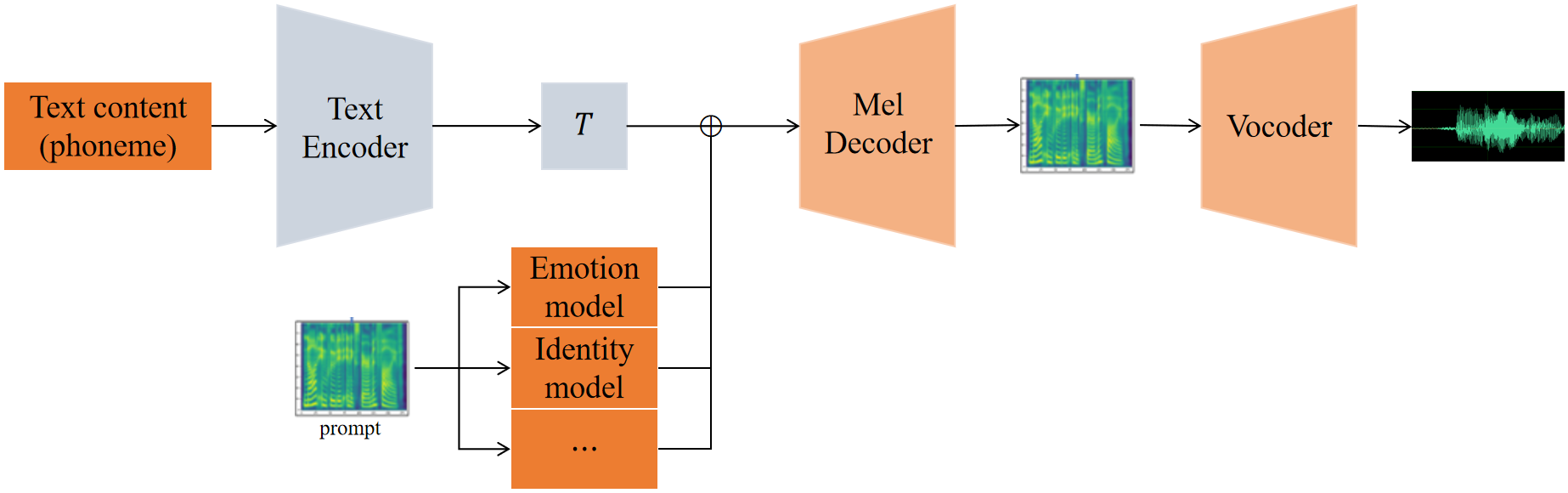}
    \caption{The general framework of Clip-TTS 2}
    \label{fig:3}
\end{figure}

I plan to launch the basic architecture of Clip-TTS 2 as shown in Figure 4. Clip-TTS is still just a basic TTS system. Although it supports multiple speakers and emotions, it is neither a zero-shot TTS nor a large model. Currently, the main research directions in the TTS field are large models and zero-shot, and an increasing number of commercial demands are also pointing towards zero-shot TTS. Therefore, my future plan is to develop Clip-TTS 2 focusing on large models and zero-shot capabilities. I aim to evolve the current version into a zero-shot TTS that can synthesize any voice timbre, emotion, prosody, etc., without the need for fine-tuning or retraining. 

\begin{figure}[ht]
    \centering
    \includegraphics[width=0.8\linewidth]{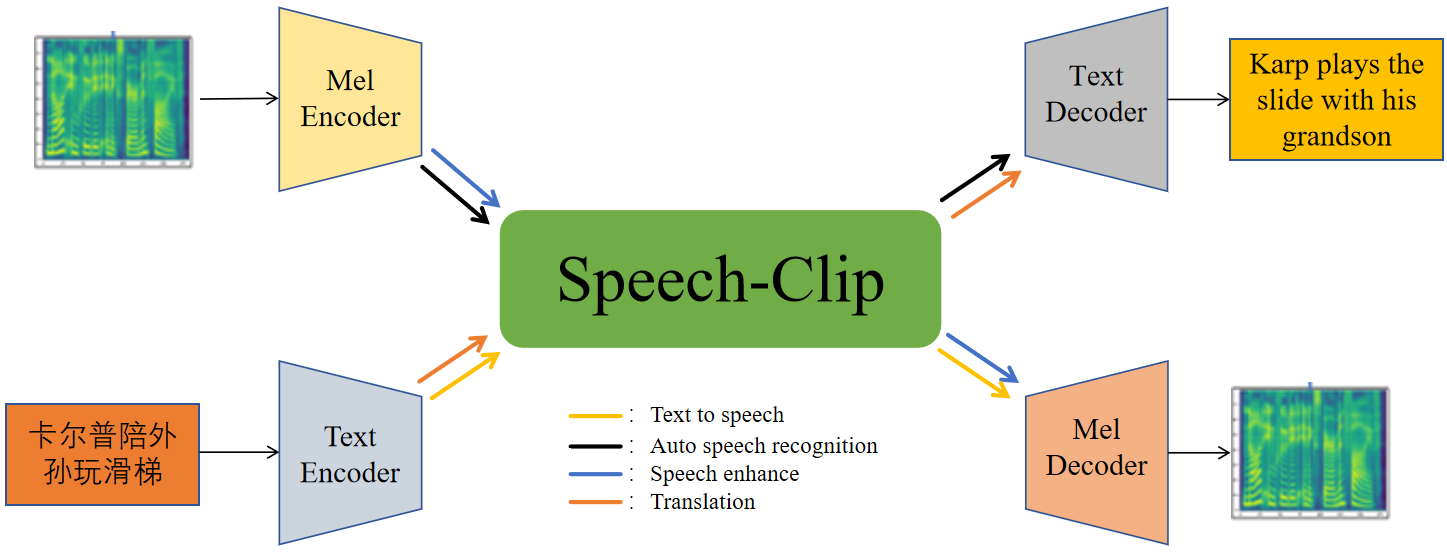}
    \caption{The downstream tasks that Speech-Clip may support}
    \label{fig:4}
\end{figure}

The potential downstream tasks supported by Speech-Clip are shown in Figure 5. In addition to TTS, Speech-Clip could perform well in three areas: speech recognition, text translation, and speech enhancement. Current mainstream speech recognition models, such as Whisper [42], take log-mel spectrograms as input and use a transformer-based encoder-decoder structure, which is very similar to my architecture. Therefore, theoretically, Speech-Clip could also achieve good results in speech recognition. Similarly, by retaining the text encoder and adding a transformer decoder, I can form a basic text translation system, which, with fine-tuning, can theoretically enable text translation functionality. For speech enhancement, I can keep the Mel encoder while adding a Mel decoder and, through fine-tuning, create a speech enhancement model. 

The above discussion is currently in the idea stage, and there is still a lot of work to be done for the actual implementation. For example, Clip-TTS 2 cannot simply achieve zero-shot by adding extra modules, nor can the downstream tasks of Speech-Clip be realized just by adding the corresponding modules. There are many details that have not yet been considered. In the future, I will work on these areas to further optimize Clip-TTS, research Clip-TTS 2, and explore the potential possibilities of Speech-Clip. 

\section{Conclusion}

In this paper, I first propose a basic Speech-Clip model, whose architecture is inspired by the Clip framework. I use contrastive learning to train the model with the aim of explicitly learning the relationship between Mel spectrograms and the corresponding text content, enabling the model to support more downstream tasks. Next, by retaining the text encoder and adding a Mel decoder, we form a basic TTS system, which I call Clip-TTS. Experimental results show that my method outperforms other approaches on the LJSpeech and AISHELL3 datasets, achieving state-of-the-art performance on the Baker dataset, and also demonstrating excellent results on multi-emotion datasets. 
Finally, I introduce and discuss the future development directions of Clip-TTS, as well as the potential downstream tasks that Speech-Clip could support, and outline the focus of future work.

\end{document}